\documentclass{llncs} 

\usepackage{amssymb}
\usepackage{textcomp}
\usepackage{amsmath}
\usepackage{comment}
\usepackage{graphicx}
\usepackage{color}
\newcommand{\Beginproof}{{\em Proof.}  }
\newcommand{\Endproof}{\hfill$\Box$\\}

\newcommand{\kobdd}{$k$-OBDD }

\newcommand{\ket}[1]{|#1\rangle}

\begin{document}

\title{Lower Bounds and Hierarchies for Quantum Memoryless Communication Protocols and Quantum Ordered Binary Decision Diagrams with Repeated Test}
\author{Farid~Ablayev$^2$\and Andris Ambainis$^{1}$ \and Kamil~Khadiev$^{1,2}$\and Aliya~Khadieva$^2$} 

\institute{Center for Quantum Computer Science, Faculty of Computing, University of Latvia, R\={\i}ga, Latvia\and Kazan Federal University, Kazan, Russia  \\ \email{fablayev@gmail.com, andris.ambainis@lu.lv, kamilhadi@gmail.com, aliyakhadi@gmail.com} 
}

\maketitle

\begin{abstract}
We explore multi-round quantum memoryless communication protocols. These are restricted version of multi-round quantum communication protocols. The ``memoryless'' term means that players forget history from previous rounds, and their behavior is obtained only by input and message from the opposite player. The model is interesting because this allows us to get lower bounds for models like automata, Ordered Binary Decision Diagrams and streaming algorithms. At the same time, we can prove stronger results with this restriction.
We present a lower bound for quantum memoryless protocols. Additionally, we show a lower bound for Disjointness function for this model.  
As an application of communication complexity results, we consider  Quantum Ordered Read-$k$-times Branching Programs ($k$-QOBDD). Our communication complexity result allows us to get lower bound for $k$-QOBDD  and to prove hierarchies for sublinear width bounded error $k$-QOBDDs, where $k=o(\sqrt{n})$. Furthermore, we prove a hierarchy for polynomial size bounded error $k$-QOBDDs for constant $k$. This result differs from the situation with an unbounded error where it is known that an increase of $k$ does not give any advantage.

\end{abstract}
\section{Introduction}

The quantum communication protocol is a well-known model. That was explored in papers \cite{rr99,Kla00,k2000,kntz07}. We consider communication ``game'' of two players: Alice and Bob. They together want to compute Boolean function. In the paper we consider the ``memoryless'' model. It means that players do not remember anything from previous rounds. So, on each round a player knows only his own part of the input and the message from an opposite player. This type of communication models was explored, for example in \cite{ckl2017,zg2015,K16}.  
On the one hand, this model is powerful enough for emulating computational models that store all information in states: automata, OBDDs, streaming algorithms, etc. On the other hand, memoryless protocol requires fewer resources and can be implemented in practice easier. Such model is useful, for example, in web applications for {\em REST} architecture.

Researchers are often interested in exploring lower bounds for computational models.
We can see different lower bounds for quantum communication models and selected functions in following papers: \cite{kntz07,ls2009,a2001,k2007,k2001}. We suggest a lower bound that demonstrates the relation between complexity characteristics of Boolean function (number of subfunctions) and complexity characteristics of the model: $N^{\pi}(f)\leq 2^{l\cdot (Ct2^{l})^2}$, where $t$ is a number of rounds, $l$ is a maximal length of a message for all rounds, $\pi$ is a partition of input variables, $N(f)$ is a number of subfunctions for a Boolean function $f$ and $C$ is some constant. Note, that a number of subfunctions is exactly  one-way deterministic communication complexity of a function. We prove this lower bound, using a technique, which was described in \cite{K16,AK13} for classical models. That based on the representation of the computational process in a linear form.  

We apply the proven lower bound to Branching programs. The model is one of well-known models of computation. That has been shown useful in a variety of domains such as hardware verification, model checking, and other applications \cite{Weg00}. It is known that the class of Boolean functions computed by polynomial size branching programs coincided with the class of functions computed by non-uniform log-space Turing machines. One of the important restrictive branching programs are oblivious read-once branching programs or Ordered Binary Decision Diagrams (OBDD) \cite{Weg00}. The OBDD model can be considered as a nonuniform automata (see, for example, \cite{ag05}). In the last decades quantum model of OBDD was considered \cite{agk01,nhk00,SS05,s06}. Researchers are interested in read-$k$-times quantum model of OBDD ($k$-QOBDD), for example \cite{hw2005}. $k$-QOBDD can be explored from automata point of view. And in that situation, we can find good algorithms for two way quantum classical automata and related models \cite{AW02,YS10B}.  

If we apply the lower bound for memoryless protocols to $k$-OBDD, then we get the relation between the characteristic of a function $f$ (a number of subfunctions, $N(f)$) and characteristics of the model: a width ($w$) and a number of layers ($k$). $N(f) \leq w^{C \cdot(kw)^2},$ for some $C=const.$ Note, that a number of subfunctions is a minimal width of a deterministic OBDD for a function \cite{Weg00}. A relation with another classical $k$-OBDDs was presented in paper \cite{K16}. A relation between deterministic OBDD and probabilistic, quantum OBDDs was presented in \cite{agkmp2005}. Furthermore, different relations between models were discussed, for example, in \cite{akv2008,agky14,g15,agky16,kk2017,gy2017,ki2017}.  
Additionally, we apply this lower bound to {\em Matrix XOR Pointer Jumping} function and present $k$-QOBDD for this function. Using this result, we prove a hierarchy of complexity classes for bounded error  $k$-QOBDDs of a sublinear width with a natural order of input variables and up to non-constant $k$. $k$-OBDD model of small width is also interesting, because, for example, the class of functions computed by constant width $poly(n)$-OBDD  equals to the well-known complexity class $NC_1$ for logarithmic depth circuits \cite{Bar89,V2008}.
For constant $k$, we apply a lower bound from communication complexity theory \cite{kntz01,kntz07} to {\em XOR Reordered Pointer Jumping} function and get a hierarchy for polynomial size $k$-QOBDD.
Recall that if we consider unbounded error $k$-OBDDs, then we have another situation. Let us consider two classes of Boolean functions: function computed by polynomial size unbounded error $k$-QOBDDs and $1$-QOBDDs. Homeister and Waack \cite{hw2005} have shown equality of these two classes. Note that due to the definition,   $k$-OBDD is polynomial width iff it is polynomial size. Similar hierarchies are known for classical cases \cite{bssw96,AK13,K16,kk2017}. But for  $k$-QOBDD it is a new result.

The paper has the following structure. Section \ref{sec:prelims} contains definitions of a communication model. In Section \ref{sec:lower}, we prove a lower bound for a bounded error quantum memoryless communication protocol and apply it to the $MXPJ_{k,p}$ function. We apply the lower bound to OBDD in Section \ref{sec:obdd}. And use these lower bounds to prove hierarchies of complexity classes for $k$-QOBDDs.

\section{Communication Model}\label{sec:prelims}
$(\pi, t, l)$ memoryless communication quantum protocol $R$ is quantum $t$-round protocol with a partition of input variables $\pi$ and a maximal length of a message $l$. On each round, a player does not remember anything about previous rounds and sends a message that depends only on an input of the player and a received message from the opposite player. Both players can measure states on any rounds, after that, they should return $1$-answer and stop computation process or continue. On the last round Player B measures qubits and answers $0$ or $1$, if someone did not do it before.  
Let us define the model in a formal way:  
\begin{definition}
Let $\pi$ be a partition of a set $X$ of variables. We define $(\pi, t, l)$ memoryless communication quantum protocol $R$ as follows: $R$ is a two party $t$-round communication protocol. Protocol $R$ uses a partition $\pi$ of variables $X$ among two quantum players Alice (A) and Bob (B). Let $\nu=(\sigma,\gamma)$ be a partition of the input $\nu$ according to $\pi$. Alice always starts the computation. All messages contain $l$ qubits.
\begin{description}
 
  \item[{\rm Round 1.}] $A$ generates the first quantum message $|m^1\rangle$ ($|m^1\rangle=|m^1\rangle(\sigma)$) and sends it to $B$.  
  \item[\rm Round 2.] $B$ generates quantum message $|m^2\rangle$ ($|m^2\rangle=|m^2\rangle(|m^1\rangle,\gamma)$), and sends it to $A$.  
  \item[\rm Round 3.] $A$ generates $|m^3\rangle$ ($|m^3\rangle=|m^3\rangle(|m^2\rangle,\sigma)$), and sends it to $B$. \item[\rm Round 4.] $B$ generates quantum message $|m^4\rangle$ ($|m^4\rangle=|m^4\rangle(|m^3\rangle,\gamma)$), and sends it to $A$. \item[...] \item[\rm Round $t$.] $B$ receives $|m^t\rangle$ and produces a result of computation $0$ or $1$, if players do not produce an answer on previous rounds. 
  \end{description}  
Both players can measure states on any rounds, after that they should return $1$-answer and stop computation process or continue.  
 The result $R(\nu)$ of computation $R$ on $\nu\in\{0,1\}^n$ is $1$ if the probability of $1$-result greats $1/2+\varepsilon$ and $R(\nu)=0$ if the probability of $1$-result less than $1/2-\varepsilon$ for some constant $\varepsilon>0$. If $Pr\{R$ returns $z\}>1/2+\varepsilon$, then $R_{\varepsilon}(\nu)=z$ , for $z\in\{0,1\}$. A Boolean function $f(X)$ is computed by protocol $R$ (presented by $R$) with bounded error if $f(\nu)=R_{\varepsilon}(\nu)$ for some $0<\varepsilon<0.5$ and for all $\nu\in\{0,1\}^{n}$. We say that protocol $R$ uses $l\cdot t$ bits communication on all rounds.

\end{definition}

\section{Lower Bounds for Communication Model}\label{sec:lower}
Let us start from the necessary definitions and notation. 

Let $\pi=(X_A,X_B)$ be a partition of the set $X$ into two sets $X_A$ and $X_B=X\backslash X_A$. Below we will use equivalent notations $f(X)$ and $f(X_A, X_B)$. Let $f|_\rho(X_B)$ be a subfunction of $f$, where $\rho$ is mapping $\rho:X_A \to \{0,1\}^{|X_A|}$ such that $\rho=\{x_{i_1}=\sigma_1,\dots,x_{i_{|X_A|}}=\sigma_{|X_A|},$ for $\{x_{i_1},\dots,x_{i_{|X_A|}}\}=X_A\}$. Function $f|_\rho(X_B)$ is obtained from $f$ by fixing values of variables from $X_A$ using values from $\rho$. Let us consider all possible subfunctions with respect to partition $\pi$: $SF^{\pi}(f)=\{f|_\rho,$ such that $ \rho:X_A \to\sigma,$ for $\sigma\in\{0,1\}^{|X_B|}\}$. Let $N^\pi(f)=|SF^{\pi}(f)|$ be the number of different subfunctions with respect to the partition $\pi$. Let the partition $half=(\{1,\dots,n/2\},\{n/2+1,\dots,n\})$
\begin{theorem}\label{th:lower-bound}
Suppose Boolean function $f(X)$ be computed by $(\pi, t, l)$ memoryless quantum communication protocol $R$ with bounded error; then we have:

$
N^{\pi}(f) \leq 2^{(1.5t+0.5 + (t-1)\log_2(2^l+2))\cdot (0.5t-0.5)(2^{l+1}+4)^2}
$
\end{theorem}

Let us describe the same result in a short way.

\begin{corollary}
Suppose Boolean function $f(X)$ be computed by $(\pi, t, l)$ memoryless quantum communication protocol $R$ with bounded error; then we have:

$
N^{\pi}(f) \leq 2^{Cl\cdot \big(t2^{l}\big)^2}
$, for some $C=const$
\end{corollary}

We present proof in  the next section.
\subsection{Proof of Theorem \ref{th:lower-bound}}

The proof of Theorem \ref{th:lower-bound} is based on a representation of a protocol's computation process in a matrix form. Then we estimate a number of special matrices, which are used for this representation.  

Now we define a sequence of matrices ${\cal M}_R(\sigma,\gamma)$ that represents a computation procedure of protocol $R$ on input $\nu= (\sigma,\gamma)$ with respect to partition $\pi$. Let $t=2k-1$, then the sequence is following:  
${\cal M}_R(\sigma,\gamma)=\Big(M_R^{(1)}(\gamma), M_R^{(1)}(\sigma), M_R^{(2)}(\gamma),$ $M_R^{(2)}(\sigma),\dots,M_R^{(k-2)}(\sigma),M_R^{(k-1)}(\gamma),M_R^{(k-1)}(\sigma),M_R^{(k)}(\gamma)\Big)$. The sequence describes a computation on rounds from $2$ to $t$.

 The $(2^{l}+2)\times (2^{l}+2)$-matrix $M_R^{(i)}(\sigma)$ describes a computation of round $2i+1$. And the $(2^{l}+2)\times (2^{l}+2)$-matrix $M_R^{(i)}(\gamma)$ describes a computation of the round $2i$.
 
Let ${\cal M}_R(\sigma)=\Big(M_R^{(1)}(\sigma),M_R^{(2)}(\sigma),\dots,M_R^{(k-2)}(\sigma),M_R^{(k-1)}(\sigma)\Big)$ be a part of sequence, which depends on $\sigma$, and ${\cal M}_R(\gamma)=\Big(M_R^{(1)}(\gamma),M_R^{(2)}(\gamma),\dots,M_R^{(k-1)}(\gamma),$ $M_R^{(k)}(\gamma)\Big)$ be a part of the sequence, which depends on $\gamma$.

Matrix $M_R^{(i)}(\gamma)$ is a complex-value matrix. It represents transformation that was made by $B$ on the round $2i$:
\begin{itemize}
 \item
Let $s=(s_1,\dots,s_{2^l+2})$ be the $r$-th row of $M_R^{(i)}(\gamma)$, for $1\leq r \leq 2^l$. Elements $(s_1,\dots,s_{2^l})$ is amplitudes for states of $l$ qubits of a message that $B$ sends on round $2i$, if he receives a message with pure state $r$. And last two elements of the row $s_{2^l+1}=s_{2^l+2}=0$.
\item
Let $s=(0,\dots,0,1,pr)$ be the  $(2^l+1)$-st row of matrix $M_R^{(i)}(\gamma)$. The row  represents a measurement event on the round $2i$. $pr$ is probability of getting $1$ on the round $2i$.  
\item
Let $s=(0,\dots,0,0,1)$ be the  $(2^l+1)$-st row of matrix $M_R^{(i)}(\gamma)$. The row represents probability of measurement on previous  rounds.
\end{itemize}

Matrices $M_R^{(i)}(\sigma)$ describe a computation of the round $2i$ and have the similar structure.  

Additionally, we define vectors $p^0_R(\sigma)$ and $q_R$, which describe the first round and accepting states after the last round, respectively. The row vector $p^0_R(\sigma)=(p_1,\dots p_{2^l+2})$ defines the message, which was formed on the first round of $R$. Each element of vector corresponds to one of $M_R^{(1)}(\gamma)$ matrix's row. $p_{2^l+1}=1$ and $p_{2^l+1}$ is the probability of $1$-result if we have measurement on the first round.

The column vector $q_R=(q_1,\dots, q_{2^l}, 0,  1)$.  Each element of vector corresponds to one of $M^{k}_R(\gamma)$ matrix's row. $q_r=1$ iff $r$ is accepting state, $q_r\in\{0,1\}$, for $1\leq r \leq 2^l$. 

Let us define $sqr$ operator that describes measurement after the last round. Let operator $sqr:{\bf C}^{2^l+2}\to{\bf R}^{2^l+2}$ be given by $sqr(z_1\dots,z_{2^l+2})=(s_1\dots,s_{2^l+2})$ , where $s_i=|z_i|^2$, for $1\leq i \leq 2^l$ and $s_i=|z_i|$ for $2^l+1\leq i \leq 2^l+2$, ${\bf C}$ is a set of complex numbers and ${\bf R}$ is a set of real numbers.

\begin{lemma}\label{linear-process}
For any input $\nu\in\{0,1\}^n$, $\nu=(\sigma, \gamma)$ we have:
\begin{equation}\label{linear-process-eq}
Pr\{R\mbox{ reaches $1$ on }\nu\}=sqr \left(p^0_R(\sigma)\left(\prod_{i=1}^{k-1}M_R^{(i)}(\gamma)M_R^{(i)}(\sigma)\right)M_R^{(k)}(\gamma) \right) \cdot q_R.
\end{equation}
\end{lemma}
\Beginproof
Let the vector $p^j=(p^j_1,\dots p^j_{2^l+2})$ be a vector that  describes the computation of $R$ after $j$ rounds on input $\nu=(\sigma,\gamma)$. Then $p^j_{r}$ for $1\leq r \leq 2^{l}$ describes amplitudes for state $r$, $p^{j}_{2^l+1}=1$ and $p^{j}_{2^l+2}$ is the probability of $1$-result if we have measurements on previous rounds and should answer $1$.  

Vector $p^j$ is computed as follows: $p^j=p^0_R(\sigma)\left(\prod_{i=1}^{\lfloor j/2 \rfloor}M_R^{(i)}(\gamma)M_R^{(i)}(\sigma)\right)$ for even $j$, and $p^j=p^0_R(\sigma)\left(\prod_{i=1}^{\lfloor j/2 \rfloor}M_R^{(i)}(\gamma)M_R^{(i)}(\sigma)\right)M_R^{(k)}(\gamma)$ for odd $j$.

By the definition of vector $q_R$ we have the following fact: $sqr \left(p^{2k-1} \right)\cdot q_R$ is the probability of reaching $1$ on input $\nu=(\sigma,\gamma)$. Hence (\ref{linear-process-eq}) is right.
\Endproof

Let us discuss the following question: ``How similar should be sequences ${\cal M}_R(\sigma, \gamma)$ and ${\cal M}_R(\sigma', \gamma)$ for equivalence of computation results for inputs $(\sigma, \gamma)$ and $(\sigma', \gamma)$?''.  
For simplifying an answer to the question, we convert complex-value matrices and vectors to real-value matrices. We use the trick from the paper \cite{MC00}. It is well known that complex numbers $c = a + bi$ can be represented by $2 \times 2$ real matrix ${\bf c}=\begin{pmatrix}a & b \\

-b & a\end{pmatrix}$
The reader can check that multiplication is faithfully reproduced and that ${\bf c}^T {\bf c} = |{\bf c}|{\bf 1}$. In the same way, a
$r \times r$ complex-value matrix can be simulated by a $2r \times 2r$ real-valued matrix. Moreover, this matrix is unitary if
the original matrix is.
Consequently, we will consider $(2^{l+1}+4)\times (2^{l+1}+4)$ real-value matrices $M^{(i)}_R(\sigma)$ and $M^{(i)}_R(\gamma)$, $(2^{l+1}+4) \times 2$ real-number matrix $p^0_R(\sigma)$ and  $2 \times (2^{l+1}+4) $ real-number matrix $q_R$. Let us pay attention to matrix $q_R=\begin{pmatrix} q_1, &  0 \\
\dots & \dots\\
 q_{2^{l+1}+4},& 0\end{pmatrix}$.
Element $q_r=1$ iff $\lceil (r+1)/2 \rceil$ is accepting state, $q_r\in\{0,1\}$, for $1\leq r \leq 2^l$. $q_{2^{l+1}+1}=q_{2^{l+1}+2}=0$ and for probability of $1$-result on previous rounds we have $q_{2^{l+1}+3}=q_{2^{l+1}+4}=1$.

Before introduction closeness of matrices, let us consider $\delta$-close metric of number equivalence.
Let $\delta \geq 0$. Two real numbers $p$ and $p'$ are called $\delta$-close if both: $-1\leq p, p'\leq 1$ and $|p-p'|<\delta$. Let $\beta \geq 0$. Two $q\times r$ matrices $M=[s_{ij}]$ and $M'=[s'_{ij}]$ are $\delta$-close iff $s_{ij}$ and $s'_{ij}$ are $\delta$-close, for any $i\in\{1,\dots, q\}$ and $j\in\{1,\dots, r\}$. We have the similar definition for vectors.

Now we can discuss an equivalence of inputs according to similarity of answer probability in the following lemma.  
\begin{lemma}\label{matrixBetaCloseProb}
Suppose inputs $(\sigma, \gamma)$ and $(\sigma',\gamma)$  such that corresponding matrices in sequences ${\cal M}_R(\sigma,\gamma)$ and ${\cal M}_R(\sigma',\gamma)$ are $\delta$-close, $p^0_R(\sigma)$ and $p^0_R(\sigma')$ are $\delta$-close; then we have: $|Pr\{\mbox{R returns $1$ on input }(\sigma,\gamma)\}-Pr\{\mbox{R returns $1$ on input }(\sigma',\gamma)\}|<2^{3k-1}(2^l+2)^{2k}\delta$, for $t=2k-1$. (See Appendix \ref{apx:betaclose})
\end{lemma}

According to above lemma, we can introduce the $\delta$-equivalence for inputs with respect to the protocol $R$. Two inputs $\sigma$ and $\sigma'$, ($\sigma,\sigma'\in\{0,1\}^{|X_A|}$) are $\delta$-equivalent if corresponding matrices in sequences ${\cal M}_R(\sigma)$ and ${\cal M}_R(\sigma')$ are $\delta$-close and $p^0_R(\sigma)$ and $p^0_R(\sigma')$ are $\delta$-close

Let us obtain possible biggest  $\delta$ such that it does not affect $1$-result probability too much.

\begin{lemma}\label{deltaeq}
Suppose inputs  $\sigma,\sigma'\in\{0,1\}^{|X_A|}$ are $\delta$-equivalent and $\delta= \varepsilon 2^{-3k}(2^l+2)^{-2k}$, then for any $\gamma\in\{0,1\}^{|X_B|}$ we have: $R_{\varepsilon}(\sigma,\gamma)=R_{\varepsilon/2}(\sigma',\gamma)$. 
\end{lemma}
\Beginproof
Let $p=Pr\{R\mbox{ reaches }1\mbox{ on }(\sigma, \gamma)\}$ and $p'=Pr\{R\mbox{ reaches }1\mbox{ on }(\sigma', \gamma)\}$. 

Probabilities $p$ and $p'$ are $2^{3k-1}(2^l+2)^{2k}\delta$-close due to Lemma \ref{matrixBetaCloseProb}. Therefore, $p$ and $p'$ are $\varepsilon/2$-close. Hence, we have:
$|p-p'|<\varepsilon/2$.
Thus, if $p>0.5+\varepsilon$ then $p'>0.5+\varepsilon/2$;
 if $p<0.5-\varepsilon$ then $p'<0.5-\varepsilon/2$.
And the claim of the lemma is right.
\Endproof

Let protocol $R$ computes Boolean function $f(X)$ with bounded error $\varepsilon$.
Let us prove that the number of subfunctions $N^\pi(f)$ is less than or equal to the number of non $\delta$-equivalent inputs $\sigma$'s with respect to the protocol $R$ and error $\varepsilon/2$, for $\delta= \varepsilon 2^{-3k}(2^l+2)^{-2k}$.
Assume that $N^\pi(f)$ greats the number of non $\delta$-equivalent $\sigma$'s. Then due to Pigeonhole principle there are two inputs $\sigma$ and $\sigma'$ and corresponding mappings $\rho$ and $\rho'$ such that $f|_\rho(X_B)\neq f|_\rho'(X_B)$, but $\sigma$ and $\sigma'$ are $\delta$-equivalent inputs. Therefore, there is $\gamma\in\{0,1\}^{|X_A|}$ such that $f|_\rho(\gamma)\neq f|_\rho'(\gamma)$, but $R_{\varepsilon/2}(\sigma,\gamma)=R_{\varepsilon/2}(\sigma',\gamma)$. This is contradiction.

If we compute the number of different non $\delta$-equivalent $\sigma$'s, we will get a claim of the lemma. Let us compute the number of different non $\delta$-equivalent $\sigma$'s. It is equal to the number of non $\delta$-close matrices from sequence ${\cal M}_R(\sigma)$ multiply the number of non $\delta$-close matrices $p^0_R(\sigma)$.  
The number of non $\delta$-close matrices in sequence ${\cal M}_R(\sigma)$ is at most
\[\left(\frac{2}{\delta} \right)^{(k-1)(2^{l+1}+4)^2}\leq
\left(\frac{2^{3k+1}(2^l+2)^{2k}}{\varepsilon}  \right)^{(k-1)(2^{l+1}+4)^2}
=\]
\[=2^{(3k+1-\log\varepsilon + 2k\log_2(2^l+2))\cdot (k-1)(2^{l+1}+4)^2}\leq\] 
\[\leq 2^{(3k+1 + 2k\log_2(2^l+2))\cdot (k-1)(2^{l+1}+4)^2}.\] 

Additionally, we have the following bound for the number of non $\delta$-close vectors $p^0(\sigma)$: 
$2^{(3k+1 + 2k\log_2(2^l+2))\cdot (2^{l+1}+4)^2}.$ Therefore, \[N^{\pi}(f)\leq 2^{(3k+1 + 2k\log_2(2^l+2))\cdot k(2^{l+1}+4)^2}=2^{(1.5t+0.5 + (t-1)\log_2(2^l+2))\cdot (0.5t-0.5)(2^{l+1}+4)^2}.\]  
\Endproof

{\bf A Lower Bound for Boolean Function $MXPJ_{k,d}$.}
Let us consider Boolean function $MXPJ_{k,d}(X)$. It is a modification of Shuffled Address Function from \cite{K15} which based on definition of Pointer Jumping ($PJ$) function from \cite{nw91,bssw96}.  

Let us present a definition of $PJ$ function for integers. Let $V_A, V_B$ be two disjoint sets (of vertexes) with $|V_A| = |V_B| = d$ and $V = V_A \cup V_B$ . Let $F_A = \{f_A : V_A \to V_B\}$, $F_B = \{f_B : V_B \to V_A\}$ and $f = (f_A, f_B):V \to V$ defined by $f(v) = f_A(v)$, if $v\in V_A$ and $f= f_B(v)$, $v\in V_B$. For each $j \geq 0$ define $f^{(j)}(v)$ by $f^{(0)}(v) = v$ , $f^{(j+1)}(v) = f(f^{(j)}(v))$. Let $v_0\in V_A$. We want to compute $g_{k,d} : F_A \times F_B \to V$ function. This is defined by $g_{k,d}(f_A, f_B) = f^{(k)}(v_0)$.    

The {\em Matrix XOR Pointer Jumping function}($MXPJ_{2k,d}$) is modification of $PJ$.
Firstly, we introduce the definition of $MatrixPJ_{2k,d}$ function. Let us consider functions $f_{A,1},\cdots f_{A,k}\in F_A$ and $f_{B,1}, \cdots f_{B,k} \in F_B$.  
On iteration $j+1$ function $f^{(j+1)}(v)=f_{j+1}(f^{(j)}(v))$, where $f_i(v)= f_{A,\lceil \frac{i}{2} \rceil}(v)$ if $i$ is odd, and $f_i(v)= f_{B,\lceil \frac{i}{2} \rceil}(v)$ otherwise.
$MatrixPJ_{2k,d}(f_{A,1},\cdots f_{A,k}, f_{B,1}, \cdots f_{B,k}) = f^{(k)}(v_0)$.
$MXPJ_{2k,d}$ is modification of $Matrix PJ_{2k,d}$. Here we take $f^{(j+1)}(v)=f_{j+1}(f^{(j)}(v)) \oplus f^{(j-1)}(v) $, for $j\geq 0$.

\begin{figure}[tbh]
\begin{center}
\includegraphics[width=5cm]{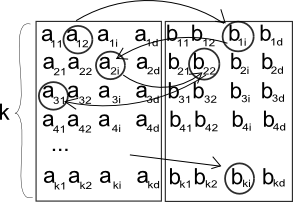}
\end{center}
\caption{Boolean function $MXPJ_{k,d}$}
\label{fig:ab}
\end{figure}

Finally, we consider a boolean version of these functions. The Boolean function $PJ_{t,n}:\{0,1\}^n\to\{0,1\}$ is $g_{k,d}$, where we encode $f_A$ in a binary string using $d\log d$ bits and do it with $f_B$ as well. The result of the function is a parity of bits from the binary representation of the result vertex's number.
For encoding functions in an input of $MXPJ_{2k,d}$, we use following order: $f_{A,1}, \dots, f_{A,k}, f_{B,1}, \dots, f_{B,k}$. Let us describe the process of computation on Figure \ref{fig:ab}. Function $f_{A,i}$ is encoded by $a_{i,1},\cdots a_{i,d}$,  and $f_{B,i}$ is encoded by $b_{i,1}, \cdots b_{i,d}$, for $i \in \{1 \cdots k\}$. We assume that $v_0=0$.  
%

Let us discuss a number of subfunctions for $MXPJ_{2k,d}$ in Lemma \ref{lm:MXPJ-subfunc} and apply our lower bound to the function in Lemma \ref{lm:MXPJ-lower-bound-cc}.  

\begin{lemma}\label{lm:MXPJ-subfunc}
For $kd\log d = o(n)$ we have:
$N^{id}(MXPJ_{2k,d})\geq  d^{\lfloor d/3-1 \rfloor (k-3)}$
\end{lemma}
\Beginproof 
The idea is similar to the proof from \cite{K15}. See Appendix \ref{apx:MXPJ-subfunc}.
\Endproof
\begin{lemma}\label{lm:MXPJ-lower-bound-cc}
$MXPJ_{2k,\lfloor \sqrt{d}\rfloor}$ cannot be computed by any (k/r,half, l) quantum memoryless communication protocol, for $C_1\sqrt{d}\log d-(C2^{2l}kl)/r^2>0$   and $C,C_1=const.$ (See Appendix \ref{apx:MXPJ-lower-bound-cc})
\end{lemma}

\section{Application to Ordered  Binary Decision Diagrams}\label{sec:obdd}
Let us start with definitions. Ordered  Read $k$-times  Branching Programs ($k$-OBDD) are a well-known model for computation of Boolean functions. For more details see \cite{Weg00}.

$k$-OBDD is a restricted version of a branching program (BP). BP over a set $X$ of $n$ Boolean variables is a directed acyclic graph with two distinguished nodes $s$ (a source node) and $t$ (a sink node). We denote it $P_{s,t}$ or just $P$. Each inner node $v$ of $P$ is associated with a variable $x\in X$. A {\em deterministic} $P$ has exactly two outgoing edges labeled $x=0$ and $x=1$ respectively for that node $v$. The program $P$ computes Boolean function $f(X)$ ($f:\{0,1\}^n \rightarrow \{0,1\}$) as follows: for each $\sigma\in\{0,1\}^n$ we let $f(\sigma)=1$ iff there exists at least one $s-t$ path (called {\em accepting} path for $\sigma$) such that all edges along this path are consistent with $\sigma$. A {\em size} of branching program $P$ is a number of nodes.
Ordered  Binary Decision Diagram (OBDD) is a BP with following restrictions: 
(i) Nodes can be partitioned into levels $V_1, \ldots, V_{\ell+1}$ such that  $s$ belongs to the first level  $V_1$ and sink node $t$ belongs to the last level $V_{\ell+1}$. Nodes from level $V_j$ have outgoing edges only to nodes of level $V_{j+1}$, for $j \le \ell$.
(ii)All inner nodes of one level are labeled by the same variable.
(iii)Each variable is tested on each path only once.

A {\em width} $w(P)$ of a program $P$ is $ w(P)=\max_{1\le j\le \ell}|V_j|. $ 
OBDD $P$ reads variables in its individual  order
$\theta(P)=(j_1,\dots,j_n)$. Let $tr_P:\{1,\dots,n\}\times\{1,\dots,w(P)\}\times\{0,1\}\to\{1,\dots,w(P)\}$ be transition function of OBDD $P$ on the level $i$.
OBDD $P$ is called {\em commutative} iff for any permutation $\theta'$ OBDD $P'$ can be constructed by reordering transition functions and $P'$ still computes the same function. Formally, $tr_{P'}(i,s,x_{\theta'(i)})=tr_{P}(\theta^{-1}(\theta'(i)),s, x_{\theta'(i)})$, for $\theta$ is the order of $P$. 
A BP $P$ is called \kobdd if it consists of $k$ layers. The  $i$-th ($1\le i\le k$) layer  $P^i$ of $P$  is  an OBDD. We call  order $\theta(P)=\theta$ the order of $P$, where $\theta(P^1)=\dots=\theta(P^k)=\theta$. $k$-OBDD $P$ is commutative iff each layer is commutative OBDD.  

Let us define a quantum $k$-OBDD ($k$-QOBDD). That is given in different terms, but you can see that they are equivalent, see \cite{agk01} for more details.
For a given $ n>0 $, a quantum OBDD $ P$ of width $w$ defined on $ \{0,1\}^n $, is a 4-tuple
$
    P=(T,\ket{\psi}_0,Accept,\pi),
$
where
$ T = \{ T_j : 1 \leq j \leq n \mbox{ and } T_j = (G_j^0,G_j^1) \} $ are ordered pairs of (left) unitary matrices representing the transitions. Here $ G_j^0 $ or $ G_j^1 $ is applied on the $j$-th step. And a choice is determined by the input bit.  
$\ket{\psi}_0$ is a initial vector from $ w $-dimensional Hilbert space over the field of complex numbers.  $ \ket{\psi}_0=\ket{q_0}$ where $ q_0 $ corresponds to the initial node.
$ Accept \subset \{1,\ldots,w\} $ is a set of accepting nodes.
$ \pi $ is a permutation of $ \{1,\ldots,n\} $ defines the order of input bits.

  For any given input $ \nu \in \{0,1\}^n $, the computation of $P$ on $\nu$ can be traced by a $ w$-dimensional vector from Hilbert space over the field of complex numbers. The initial one is $ \ket{\psi}_0$. In each step $j$, $1 \leq j \leq n$, the input bit $ x_{\theta(j)} $ is tested and then the corresponding unitary operator is applied:  
$
    \ket{\psi}_j = G_j^{x_{\theta(j)}} (\ket{\psi}_{j-1}),
$ 
where  $ \ket{\psi}_j $ represents the state of the system after the $ j$-th step, for $ 1 \leq j \leq n $.
We can measure one of qubits. Let the program was in state $\ket{\psi}=(v_1, \dots, v_w)$ before measurement and let us measure the $i$-th qubit. And let states with numbers $j^0_1,\dots,j^0_{w/2}$ correspond to $0$ value of the $i$-th qubit, and states with numbers $j^1_1,\dots,j^1_{w/2}$ correspond to $1$ value of the $i$-th qubit.
The result of measurement of $i$-th qubit is $1$ with probability $pr_1= \sum_{u=1}^{w/2}|v_{j^1_u}|^2$ and $0$ with probability $pr_0=1-pr_1$.
In the end of computation program $P$ measures all qubits. The accepting (return $1$) probability $Pr_{accept}(\sigma)$ of $ P_n $ on input $ \sigma $ is 
$
    Pr_{accept}(\nu)=\sum_{i \in Accept} v^2_i.
$, for $ \ket{\psi}_n=(v_1,\dots,v_w)$.

Let $P_{\varepsilon}(\nu)=1$ if $P$ accepts input $\nu\in\{0,1\}^n$ with  probability at least $ 0.5+\varepsilon$, and $P_{\varepsilon}(\nu)=0$ if  $P$ accepts input $\nu\in\{0,1\}^n$ with  probability at most $ 0.5-\varepsilon$, for $ \varepsilon \in (0,0.5] $.
We say that a function $f$ is computed by $ P$ with bounded error if there exists an $ \varepsilon \in (0,0.5] $ such that $ P_{\varepsilon}(\nu)=f(\nu)$ for any  $\nu\in\{0,1\}^n$. We can say that $P$ computes $f$ with bounded error $0.5-\varepsilon$. 

Quantum $k$-OBDD ($k$-QOBDD) is Quantum Branching program with $k$ layers. Each layer is QOBDD, and each layer has the same order $\theta$. We allow measurement for $k$-QOBDD during the computation, but after that, it should stop and accept an input or continue the computation.  $k$-$id$-QOBDD is $k$-QOBDD with the natural order of input bits $id=(1,\dots,n)$.  

Let $k$-{\bf QOBDD}$_{{\cal W}}$ be a set of Boolean functions that can be computed by bounded error $k$-QOBDDs of width $w$, for $w\in{\cal W}$. $k$-$id$-{\bf QOBDD}$_{{\cal W}}$ is the same for bounded error $k$-$id$-QOBDDs.  
As ${\cal W}$ we will consider only ``good'' sets $G_{k,r}$, for integer $k=k(n)$, $r=r(n)$. The set ${\cal W}$ belongs to $G_{k,r}$ if this is ths set of integers with following properties:
(i) if $w\in {\cal W}$, then $\lfloor \sqrt{w}\rfloor, \lfloor \sqrt{w}\rfloor^2\in {\cal W}$;
(ii) $k^2w^2\log w = o(n)$, for any $w\in {\cal W}$;
(ii) $C_1\sqrt{w}\log w-(Cv^2k\log v)/r^2>0$  for any $w,v\in {\cal W}$ and $C,C_1=const.$
%
Let BQP$_{\varepsilon}$-$k$QOBDD be a set of Boolean functions that can be computed by polynomial size $k$-QOBDDs with probability of a right answer at most $1-\varepsilon$ or an error at least $\varepsilon$. We can consider similar classes for deterministic model (P-$k$OBDD) and bounded error probabilistic model (BP$_{\varepsilon}$-$k$OBDD)

{\bf Lower Bound for Ordered  Binary Decision Diagrams.}
Let us start from necessary definitions and notation. Let $\Theta(n)$ be the set of all permutations of $\{1,\dots,n\}$.
Let the partition $\pi(\theta,u)=(X_A,X_B)=(\{x_{j_{1}},\dots,x_{j_u}\},$ $\{x_{j_{u+1}},\dots,x_{j_n}\})$, for the  permutation $\theta=(j_1,\dots, j_n)\in \Theta(n), 1<u<n$. We denote $\Pi(\theta)=\{\pi(\theta,u): 1<u<n\}$.
Let $ N^\theta(f)=   \max_{\pi\in \Pi(\theta)} N^\pi(f),
N(f)=\min_{\theta\in \Theta(n)}N^\theta(f). $ 

We can emulate $k$-QOBDD $P$ of width $w$ and order $\theta$ with $(\pi,t,l)$ memoryless communication quantum protocol $R$, such that $\pi\in \Pi(\theta)$, $t=2k-1$ and $2^l=w$. Such emulation is described, for example, in \cite{K16}. Therefore, the lower bound for $k$-QOBDD follows from Theorem \ref{th:lower-bound}.

\begin{theorem}\label{th:kqobdd-lower-bound}
Suppose function $f(X)$ is computed  by bounded error $k$-QOBDD $P$ of  width $w$;
then $N(f) \leq 2^{d}, \mbox{ for }d=\big(3k+1 + 2k\log_2(w+2)\big)\cdot k(2w+4)^2.$
\end{theorem}
\begin{corollary}\label{cr:kqobdd-lower-bound-const}
Suppose function $f(X)$ is computed  by bounded error $k$-QOBDD $P$ of  width $w$;
then $N(f) \leq w^{C \cdot(kw)^2}, \mbox{ for some }C=const. $
\end{corollary}

Note that this lower bound gives us relation with deterministic OBDD complexity of function, because $N(f)$ is the width of better deterministic OBDD for function \cite{Weg00}.
Let us apply this lower bound to  $MXPJ_{k,d}(X)$ function. 
\begin{lemma}\label{lm:MXPJ-lower-bound}
Let ${\cal W}\in G_{k,r}$, for integers $k=k(n),r=r(n)$. Then

$MXPJ_{2k,\lfloor \sqrt{d}\rfloor}\not\in \lfloor k/r\rfloor$-id-{\bf QOBDD}$_{{\cal W}}$.(See Appendix \ref{apx:MXPJ-lower})
\end{lemma}

\section{Hierarchy Results}\label{sec:hierarchy}
{\bf Hierarchy for Sublinear Width.}$\quad$
Firstly, let us discuss upper bound for $MXPJ_{k,d}$ function. The Proof is in Appendix \ref{apx:MXPJ-upper}
\begin{lemma}\label{lm:MXPJ-upper-bound}
There is exact $k$-id-QOBDD $P$ of width $d^2$ which computes  $MXPJ_{2k,d}$. 
\end{lemma}
Using above lemma and lower bound from Lemma \ref{lm:MXPJ-lower-bound}, we get hierarchy results.
\begin{theorem}\label{th:hierarchy-small}
Suppose ${\cal W}\in G_{k,r}$, for integers $k=k(n),r=r(n)$, then: 

$\lfloor k/r\rfloor$-id-{\bf QOBDD}$_{{\cal W}}\subsetneq k$-id-{\bf QOBDD}$_{{\cal W}}$. (See Appendix \ref{apx:hierarchy-small})
\end{theorem}
Partial cases are hierarchies for the following classes: $k$-id-{\bf QOBDD}$_{CONST}$, $k$-id-{\bf QOBDD}$_{PLOG}$ and $k$-id-{\bf QOBDD}$_{SUBLIN(\alpha)}$. Here $CONST=\{w:w=const\}$, $PLOG=\{w:w=(\log n)^{O(1)}\}$, $SUBLIN(\alpha)=\{w:w=O(n^\alpha),$ for $0<\alpha<1\}$.
\begin{corollary}\label{cr:from-hierarchy-small}
Claim 1.
$\lfloor \sqrt{k}/r\rfloor$-id-{\bf QOBDD}$_{CONST}\subsetneq k$-id-{\bf QOBDD}$_{CONST}$, for $k = o(\sqrt{n})$, $\sqrt{k}>r$, $1=o(r)$.

Claim 2.
$\lfloor \sqrt{k}/n^{r}\rfloor$-id-{\bf QOBDD}$_{PLOG}\subsetneq k$-id-{\bf QOBDD}$_{PLOG}$, for $k = o(n^{0.5-\delta})$, $\sqrt{k}>n^r$, $r>0, \delta>0$.

Claim 3.
$\lfloor \sqrt{k}/n^{\alpha+r}\rfloor$-id-{\bf QOBDD}$_{SUBLIN(\alpha)}\subsetneq k$-id-{\bf QOBDD}$_{SUBLIN(\alpha)}$, for $k = o(n^{0.5-\alpha-\delta})$, $\sqrt{k}>n^{\alpha+r}$, $r>0, \delta>0$ and $0>\alpha>1/6-\delta/3-2r/3$. 
\end{corollary}
\Beginproof

Let us consider Claim 1. We get conditions 1 and 2 of $G_{k,r}$, because ${\cal W}=CONST, k=o(\sqrt{n})$. Let us consider condition 3 and $r'=\sqrt{k}r$. Then $C_1w\log w-(Cv^2k\log v)/r'^2=C' -C''/r^2>0$ for $C',C''=const$, because $1=o(r)$. Therefore, due to Theorem \ref{th:hierarchy-small}, we have:

$\lfloor k/r'\rfloor\mbox{-id-{\bf QOBDD}}_{{CONST}}\subsetneq k\mbox{-id-{\bf QOBDD}}_{CONST}$ and we get Claim 1.

Let us consider Claim 2. We get conditions 1 and 2 of $G_{k,r}$, because ${\cal W}=PLOG, k=o(n^{0.5-\delta})$. Let us consider condition 3 and $r'=\sqrt{k}n^r$. Then $C_1w\log w-(Cv^2k\log v)/r'^2>C' -C''O(n^{r})/n^{2r}=C'-C''/n^{r}>0$ for $C',C''=const$. Therefore, due to Theorem \ref{th:hierarchy-small}, we have:

$\lfloor k/r'\rfloor\mbox{-id-{\bf QOBDD}}_{{PLOG}}\subsetneq k\mbox{-id-{\bf QOBDD}}_{PLOG}$ and we get Claim 2.  

Let us consider Claim 3. We get conditions 1 and 2 of $G_{k,r}$, because ${\cal W}=SUBLIN(\alpha), k = o(n^{0.5-\alpha-\delta})$, $\sqrt{k}>n^{\alpha+r}$, $r>0, \delta>0$
 and $0>\alpha>1/6-\delta/3-2r/3$. Let us consider condition
 3 and $r'=\sqrt{k}n^{\alpha+r}$. Then $C_1w\log w-(Cv^2k\log v)/r'^2>C' -C''O(n^{2\alpha+r})/n^{2\alpha+2r}=C'-C''/n^{r}>0$ for $C',C''=const$. 
Therefore, due to Theorem \ref{th:hierarchy-small}, we have:

$\lfloor k/r'\rfloor\mbox{-id-{\bf QOBDD}}_{{SUBLIN(\alpha)}}\subsetneq k\mbox{-id-{\bf QOBDD}}_{SUBLIN(\alpha)}$
and we get Claim 3.
\Endproof
{\bf Hierarchy for Polynomial Size.}$\quad$
Let us consider a Boolean function $XRPJ_{k,n}$, it is a modification of boolean version of $PJ_{k,n}$ function using reordering method from \cite{kk2017}. We add address for each bit of input and compute with respect to the address in original input. If we meet bits with the same address, then we consider their XOR.
$XRPJ_{k,n}$ is a total version of xor-reordered $PJ_{k,n}$, details in \cite{kk2017}. Let us define this formally.

Let us split the input $X=(x_1,\dots,x_n)$ to $b$ blocks with $n/b$ elements, such that $b\lceil \log_2 b + 1\rceil=n$, therefore, $b=O(n/\log n)$. And let $Adr(X,i)$ be an integer such that its binary representation is first $\lceil \log_2 b\rceil$ bits of the $i$-th block. Let $Val(X,i)$ be a value of the bit number $\lceil \log_2b+1 \rceil$ from the block $i$, for $i\in\{0,\dots,b-1\}$. Let $2d\lceil\log d\rceil=b$ and $V_A=\{0,\dots,d-1\}$, $V_B=\{d,\dots,2d-1\}$. Hence, $d = O(n/\log^2n)$.
Let function $BV:\{0,1\}^n\times\{0,\dots, 2d-1\}\to \{0,\dots, d-1\}$ be the following:

\[BV(X,v) = \sum_{j=(v-1)\log b + 1}^{v\log b}2^{j - (v-1)\log b-1}\cdot \bigoplus_{i: Adr(X,i)=j} Val(X,i)\]

Then $f_A(v)=BV(X,v)+d$, $f_B(v)=BV(X,v)$.
Let $r=g_{t,a}(f_A,f_B)$, then 
\[XRPJ_{t,n}(X)=\bigoplus_{i:(r-1)\log b < Adr(X,i)\leq r\log b}Val(X,i)\].

Let us prove lower and upper bounds for $XRPJ_{k,n}$:
\begin{lemma}\label{lm:pj-kobdd}
Claim 1. Suppose  $k$-QOBDD $P$ of width $w$ computes $XRPJ_{2k-1,n}(X)$ with bounded error at least $1/8$; then $w\geq 2^{r}$, for $r=n/(k2^{O(k)})-k\log n$. 

Claim 2. There is exact $2k$-QOBDD $P$ of width $O(n^{2k+1})$ computing $XRPJ_{2k-1,n}(X)$. 
\end{lemma}
\Beginproof
The proof of the first claim is based on lower bound for quantum communication complexity from \cite{kntz01,kntz07}. We apply the bound in the similar way as in \cite{kk2017}.

 Assume that $XRPJ_{2k-1,n}$ is computed by $k$-QOBDD $P$ of width $w=2^{o(r)}$. $k$-QOBDD $P$ can be simulated by $2k-1$-round quantum communication protocol $R$, which sends at most $\lceil\log_2w\rceil(2k-1)$ bits. For prove this fact look, for example, at \cite{K16}. Let us consider only inputs from the set $\Sigma\subset\{0,1\}^n$ such that for $\sigma\in\Sigma$ we have $Adr(\sigma,i)=i+b$, for $0\leq i\leq b-1$ and $Adr(\sigma,i)=i-b$, for $b\leq i\leq 2b-1$, $b \log b = n$. For these inputs, our protocol will just compute $PJ_{2k-1,b}$, but in a communication game $B$ starts the computation. Therefore, from the protocol $R$ we can get the protocol $R'$ such that $B$ starts computation. The protocol $R'$ computes $PJ_{2k-1,b}$ and sends at most $\lceil\log_2w\rceil(2k-1)$ bits. It means $Q^{B,2k-1}_{1/8}(PJ_{2k-1,b})=o(r)$. This contradicts with results from  quantum communication complexity \cite{kntz01,kntz07}.

For the proof of the second claim, we  construct $2k$-id-QOBDD for the function. The main idea is to store a pointer for current steps and use new qubits for a new step. And we apply reordering method from \cite{kk2017}. See Appendix \ref{apx:pj-kobdd} for the full proof.
\Endproof

Using this lemma, we can prove the following hierarchy result:  
\begin{theorem}\label{th:hierarchy-big}
BQP$_{1/8}$-$k$OBDD$\subsetneq$BQP$_{1/8}$-$(2k)$QOBDD, for $k>0,k=const$. (See Appendix \ref{apx:hierarchy-big})
\end{theorem}

Both hierarchies from Corollary \ref{cr:from-hierarchy-small} and Theorem \ref{th:hierarchy-big} are interesting, because we cannot apply lower bound from Theorem \ref{th:kqobdd-lower-bound} to polynomial width, at the same time, we cannot use results from Lemma \ref{lm:pj-kobdd} to sublinear width.

{\bf Acknowledgements.} 
The work is partially supported by ERC Advanced Grant MQC.
The work is performed according to the Russian Government Program of Competitive Growth of Kazan Federal University.

\bibliographystyle{splncs03}
\bibliography{tcs}
\newpage
\appendix
\section{The Proof of Lemma \ref{matrixBetaCloseProb}}\label{apx:betaclose}
Before the proof of the lemma, let us discuss some properties of a $\delta$-close metric.

\begin{property}\label{prop:mera_prod}
 If $a$ and $b$ are $\delta$-close; $c$ and $d$ are $\delta$-close; $-1\leq a,b,c,d \leq 1$,  then $ac$ and $bd$ are $2\delta$-close. 
\end{property}

 \Beginproof
Let us estimate the statement $|ac-bd|$:

\[|ac-bd|=|ac-bc+bc-bd|\leq |a-b||c|+|b||c-d|\leq \delta|c| +|b| \delta\leq 2\delta \]
 \Endproof
 
 \begin{property}\label{prop:mera_const_prod}
 If $a$ and $b$ are $\delta$-close; $-1\leq a,b,c \leq 1$,  then $ac$ and $bc$ are $\delta$-close. 
\end{property}

 \Beginproof
Let us estimate the statement $|ac-bc|$:

\[|ac-bc|=|c||a-b|\leq \delta|c| \leq \delta \]
 \Endproof
 
 \begin{property}\label{prop:mera_inner_prod}
If vectors with $r$ elements $a$ and $b$ are $\delta$-close; vectors with $r$ elements $c$ and $d$ are $\delta$-close; $-1\leq a_i,b_i,c_i,d_i \leq 1$ for $i\in\{1,\dots,r\}$, then inner products $a\cdot c$ and $b\cdot d$ are $(2r\delta)$-close.
\end{property}
 \Beginproof
Let us consider the difference of inner products:
 \[
|a\cdot c-b\cdot d|=\left|\sum_{i=1}^{r}a_ic_i-\sum_{i=1}^{r}b_id_i \right|=\left|\sum_{i=1}^{r}(a_ic_i-b_id_i)\right|\leq
\sum_{i=1}^{r}|a_ic_i-b_id_i|
\]

Due to Property \ref{prop:mera_prod}, we have: $|a\cdot c-b\cdot d|<2r\delta$.
\Endproof
 \begin{property}\label{prop:mera_const_inner_prod}
If vectors with $r$ elements $a$ and $b$ are $\delta$-close; $c$ is a vector with $r$ elements; $-1\leq a_i,b_i,c_i \leq 1$, for $i\in\{1,\dots,r\}$, then inner products $a\cdot c$ and $b\cdot c$ are $(r\delta)$-close.  
\end{property}
 \Beginproof
Let us consider the difference of inner products:
 \[
|a\cdot c-b\cdot c|=\left|\sum_{i=1}^{r}a_ic_i-\sum_{i=1}^{r}b_ic_i \right|=\left|\sum_{i=1}^{r}(a_ic_i-b_ic_i)\right|\leq
\sum_{i=1}^{r}|a_ic_i-b_ic_i|
\]

Due to Property \ref{prop:mera_const_prod}, we have: $|a\cdot c-b\cdot d|<r\delta$.
\Endproof

\begin{property}\label{prop:mera_mul_vec_mtrx}
 If $q\times r$-matrices $A$ and $B$ are $\delta$-close; $r\times z$-matrices $D$ and $E$ are $\delta$-close such that all elements of these matrices at most $1$ by absolute value, then $q \times z$-matrices $AD$ and $BE$ are $2r\delta$-close.
\end{property}
 \Beginproof
Let $d_1,\dots,d_z$ are columns of $D$; $e_1,\dots, e_z$ are columns of $E$, then $d_i$ and $e_i$, for $i\in\{1,\dots,z\}$, are $\delta$-close. Let $a_1,\dots,a_q$ are rows of $A$; $b_1,\dots, b_q$ are rows of $B$, then $a_i$ and $b_i$, for $i\in\{1,\dots,q\}$, are $\delta$-close due to definition.

Therefore, $G=AD$ and $H=BE$ are $2r\delta$-close, because elements of matrixes $g_{ij}=a_i \cdot d_j$, $h_{ij}=b_i \cdot e_j$ and $a_i \cdot d_j$ and $b_i \cdot e_j$ are $2r\delta$-close due to Property \ref{prop:mera_inner_prod}.  
 \Endproof
 
 \begin{property}\label{prop:mera_const_mul_vec_mtrx}
 If $q\times r$-matrices $A$ and $B$ are $\delta$-close; $D$ is $r\times z$-matrix, such that all elements of these matrices at most $1$ by absolute value, then $q \times z$-matrices $AD$ and $BD$ are $r\delta$-close.
\end{property}
 \Beginproof
Let $d_1,\dots,d_z$ are columns of $D$. Let $a_1,\dots,a_q$ are rows of $A$; $b_1,\dots, b_q$ are rows of $B$, then $a_i$ and $b_i$, for $i\in\{1,\dots,q\}$, are $\delta$-close due to definition.

Therefore, $G=AD$ and $H=BD$ are $r\delta$-close, because elements of matrixes $g_{ij}=a_i \cdot d_j$, $h_{ij}=b_i \cdot d_j$ and $a_i \cdot d_j$ and $b_i \cdot d_j$ are $r\delta$-close due to Property \ref{prop:mera_const_inner_prod}.
 \Endproof

And now let us prove Lemma \ref{matrixBetaCloseProb}.


 Firstly, matrices $p^0_R(\sigma)\left(\prod_{i=1}^{k-1}M_R^{(i)}(\gamma)M_R^{(i)}(\sigma)\right)M_R^{(k)}(\gamma)$ and 
 
 $p^0_R(\sigma')\left(\prod_{i=1}^{k-1}M_R^{(i)}(\gamma)M_R^{(i)}(\sigma')\right)M_R^{(k)}(\gamma)$ are $2^{k-1}(2(2^l+2))^{2k-1}\delta$-close  due to Properties \ref{prop:mera_mul_vec_mtrx} and \ref{prop:mera_const_mul_vec_mtrx}.
 
 Secondly, let us consider matrices $ sqr \left(p^0_R(\sigma)\left(\prod_{i=1}^{k-1}M_R^{(i)}(\gamma)M_R^{(i)}(\sigma)\right)M_R^{(k)}(\gamma) \right)$ and $sqr \left(p^0_R(\sigma')\left(\prod_{i=1}^{k-1}M_R^{(i)}(\gamma)M_R^{(i)}(\sigma')\right)M_R^{(k)}(\gamma) \right)$. These matrices are $
 2^{k}(2(2^l+2))^{2k-1}\delta$-close due to  the following fact: if $|a-b|<c$ then $|a^2-b^2|=|a-b|(|a|+|b|)<2|a-b|<2c$.

 Finally, $Pr\{R\mbox{ reaches }1\mbox{ on }(\sigma, \gamma)\}=sqr \left(p^0_R(\sigma)\left(\prod_{i=1}^{k-1}M_R^{(i)}(\gamma)M_R^{(i)}(\sigma)\right)M_R^{(k)}(\gamma) \right)\cdot q(\gamma)$ and $Pr\{R\mbox{ reaches }1\mbox{ on }(\sigma', \gamma)\}=sqr \left(p^0_R(\sigma')\left(\prod_{i=1}^{k-1}M_R^{(i)}(\gamma)M_R^{(i)}(\sigma')\right)M_R^{(k)}(\gamma) \right)\cdot q(\gamma)$ are $2^{3k-1}(2^l+2)^{2k}\delta$-close due to previous fact and Property \ref{prop:mera_const_inner_prod}.
 \Endproof


\section{The Proof of Lemma \ref{lm:MXPJ-lower-bound-cc}}\label{apx:MXPJ-lower-bound-cc}

Let us compute the following rate for $v= 2^l$.

\[\frac{N^{id}(MXPJ_{2k,\lfloor \sqrt{d} \rfloor})}{v^{C\cdot (vk/r)^2}}
\geq \frac{d^{\lfloor \lfloor \sqrt{d} \rfloor/3-1 \rfloor (k-3)}}{v^{C\cdot (vk/r)^2}}
\geq \frac{d^{k\sqrt{d}/48}}{v^{C\cdot (vk/r)^2}}=\]
\[=2^{C_1k\sqrt{d}\log d-(Cv^2k^2\log v)/r^2}=2^{k(C_1\sqrt{d}\log d-(Cv^2k\log v)/r^2)}>1
\]
for $C_1=const$.

Hence $N(MXPJ_{2k,\lfloor \sqrt{d} \rfloor}) > 2^{lC\cdot (2^lk/r)^2}$ for any $s\in {\cal L}$. And by Theorem \ref{th:lower-bound} we get the claim of the lemma.
\section{The Proof of Lemma \ref{lm:MXPJ-lower-bound}}\label{apx:MXPJ-lower}

Let us compute the following rate for $v\in {\cal W}$.

\[\frac{N^{id}(MXPJ_{2k,\lfloor \sqrt{d} \rfloor})}{v^{C\cdot (vk/r)^2}}
\geq \frac{d^{\lfloor \lfloor \sqrt{d} \rfloor/3-1 \rfloor (k-3)}}{v^{C\cdot (vk/r)^2}}
\geq \frac{d^{k\sqrt{d}/48}}{v^{C\cdot (vk/r)^2}}=\]
\[=2^{C_1k\sqrt{d}\log d-(Cv^2k^2\log v)/r^2}=2^{k(C_1\sqrt{d}\log d-(Cv^2k\log v)/r^2)}>1
\]
for $C_1=const$.

Hence $N(MXPJ_{2k,\lfloor \sqrt{d} \rfloor}) > v^{C\cdot (vk/r)^2}$ for any $v\in {\cal W}$. And by Theorem \ref{th:kqobdd-lower-bound} we have  $MXPJ_{2k,\lfloor \sqrt{d} \rfloor}\not\in \lfloor k/r\rfloor$-{\bf QOBDD}$_{{\cal W}}$

\section{The Proof of Lemma \ref{lm:MXPJ-upper-bound}}\label{apx:MXPJ-upper}
 
To construct zero-error (exact) $k$-QOBDD $P$ of width $d^2$ for the Boolean function $MXPJ_{2k,d}$ we take two quantum registers of size $t=\lceil \log_2 {d} \rceil$. We denote them $| \phi \rangle$ and $| \psi\rangle$. Initial state of quantum system is  $| \phi \rangle \otimes| \psi\rangle=|0 \dots 0\rangle$.

By the definition of function $MXPJ_{2k,d}$, input is separated into $2dk$ blocks by $t=\lceil \log_2 {d} \rceil$ bits. Blocks encode integers $a_{i1},a_{i2} \cdots a_{id}$ for $i \in \{1,\cdots k\}$ in the first part of input; and $b_{i1},b_{i2} \cdots b_{id}$ for $i \in \{1,\cdots k\}$ in the second part (see Figure \ref{fig:ab}). Let elements of the block representing $a_{ij}$ be $X^{i,j}=(x^{i,j}_0, \dots, x^{i,j}_{t-1})$ for $i \in \{1,\cdots k\}, j \in \{1,\cdots d\}$ and elements of the block representing $b_{ij}$ be $Y^{i,j}=(y^{i,j}_0, \dots, y^{i,j}_{t-1})$ for $i \in \{1,\cdots k\}, j \in \{1,\cdots d\}$

\begin{figure}[tbh]
\begin{center}
\includegraphics[width=5cm]{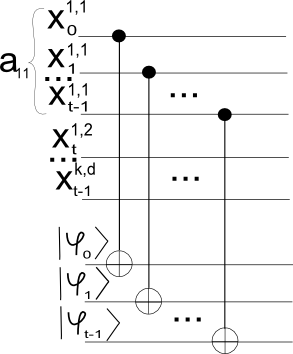}
\end{center}
\caption{Storing bits into quantum register}
\label{fig:sch1}
\end{figure}

On the first layer the program $P$ reads input bits $x^{1,1}_0,\cdots x^{1,1}_{t-1}$ one by one and stores them in $| \phi \rangle$. The transformations for this procedure are described on Figure \ref{fig:sch1}. Here we use quantum circuits notation. You can see more detailed description in \cite{akv2008}. The program applies XOR gate (or CNOT gate) to qubit. It means the following: if $x=0$ then $P$ applies $I$ gate to the corresponding qubit, and $NOT$ gate otherwise.
$I = 
\begin{pmatrix}
1 & 0 \\
0 & 1 
\end{pmatrix}$, 
 $NOT = 
\begin{pmatrix}
0 & 1 \\
1 & 0 
\end{pmatrix}$

The program reads input bits of the first part corresponding to all blocks  $a_{ij}$ except $a_{11}$  without any modification of qubits.

 Thus, the vector $| \phi \rangle$ corresponds to the address of a value from the second part.
 
\begin{figure}[tbh]
\begin{center}
\includegraphics[width=8cm]{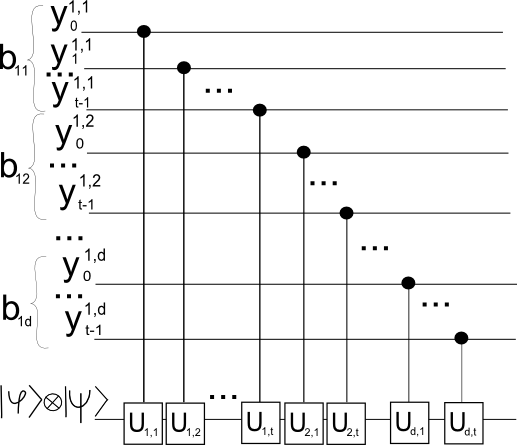}
\end{center}
\caption{Storing next blocks.}
\label{fig:sch2}
\end{figure}

 Then $k$-QOBDD $P$  reads input bits for $b_{11}, \dots, b_{1d}$ blocks and  modifies   $| \phi \rangle| \times |\psi \rangle $ as presented in Figure \ref{fig:sch2}.
 
Here we apply gate $ U_{i,j}$ to all qubits. The gate is presented by a unitary matrices pair $(U^0_{i,j}, U^1_{i,j})$.
\begin{center}
$U^p_{i,j} = 
\begin{pmatrix}
I^{\otimes t} & \cdots &0 & 0 & 0 & \cdots & 0\\         
\vdots  & \ddots&\vdots & \vdots  & \vdots & \ddots  & \vdots\\
0 & \cdots &I^{\otimes t}& 0  & 0 & \cdots & 0\\
0 & \cdots &0& R^p_j & 0 & \cdots & 0\\
0 & \cdots &0& 0  &  I^{\otimes t} & \cdots & 0\\
 \vdots & \ddots& \vdots  & \vdots & \vdots & \ddots  & \vdots\\
0 & \cdots &0& 0  & 0 & \cdots & I^{\otimes t}
\end{pmatrix}$, for $p\in\{0,1\}$
\end{center}
$d^2 \times d^2$ matrix $U^p_{i,j}$  is block-diagonal such that   $d \times d$ matrix $R_j$ in the $i$-th block. $I^{\otimes t}$ is a unit matrix of size $d \times d$.

The $d \times d$ matrix $R^p_1=XOR^p \otimes I \otimes \cdots \otimes I$, for $XOR^0=I$ and $XOR^1=NOT$ as presented above.

The $d \times d$ matrix $R^p_j=I \otimes \dots  \otimes I \otimes XOR^p \otimes I \otimes \cdots \otimes I$, where matrix $XOR^p$ is in the $j$-th position of this sequence. It means that the program applies $XOR$-gate to the $j$-th qubit of $|\psi\rangle$.   

On the $r$-th layer the program $P$ applies transformations only on the blocks $X^{r,j}$ and $Y^{r,j}$ for $j\in\{1,\dots,d\}$. These transformations are similar to transformations from the schema on Figure \ref{fig:sch2}. The difference is  following: if the program reads bits from the block $X^{r,j}$ then $P$ transforms system  $| \psi \rangle \times |\phi \rangle $. And if $P$ reads bits from the block $Y^{r,j}$, then the program applies transformations to the $| \phi \rangle\times |\psi \rangle $.

At the end, $k$-QOBDD $P$ measures qubits $|\psi \rangle$ and gets the number $f^{2k}(0)$ as a result of the computation.  If the number of ones in binary representation for a number of state is odd, then we mark it as accepting state. And if the number is even then we mark it as rejecting state. Therefore, $P$ computes $MXPJ_{2k,d}$ with probability $1$.

\section{Proof of Theorem \ref{th:hierarchy-small}}\label{apx:hierarchy-small}

Let us take $d\in {\cal W}$, then $\lfloor \sqrt{d}\rfloor\in {\cal W}, \lfloor \sqrt{d}\rfloor^2\in {\cal W}$. Due to Lemma \ref{lm:MXPJ-upper-bound}, we know that $MXPJ_{2k,\lfloor \sqrt{d}\rfloor}\in k$-id-{\bf QOBDD}$_{\cal W}$. At the same time, $MXPJ_{2k,\lfloor \sqrt{d}\rfloor}\not\in \lfloor k/r \rfloor$-id-{\bf QOBDD}$_{\cal W}$, because of Lemma \ref{lm:MXPJ-lower-bound}. So it means that, $k$-id-{\bf QOBDD}$_{\cal W}\subsetneq \lfloor k/r \rfloor$-id-{\bf QOBDD}$_{\cal W}$.  

\section{The Proof of Upper Bound from Lemma \ref{lm:pj-kobdd}}\label{apx:pj-kobdd}
Let us prove the second claim:
there is exact $2k$-QOBDD $P'$ of width $O(n^{2k})$ computing $XRPJ_{2k-1,n}(X)$. 

Firstly, let us construct a commutative $2k$-id-QOBDD $P$ of width $O(n^{2k})$ for $PJ_{2k-1,n}$ and then using the result from \cite{kk2017} we get $2k$-QOBDD $P'$ of width $O(n^{2k+1})$ for a xor-reordered version of $PJ_{2k-1,n}$, (or $XRPJ_{2k-1,n}(X)$).  

Let us construct $2k$-id-QOBDD $P$. The program has $2k+1$ groups of qubits $|\psi^0\rangle,|\psi^1\rangle,\dots,|\psi^{2k}\rangle$, each one contains $\lceil \log d \rceil+1$ qubits, except the $2k$-th one. The last one contains only one qubit. The initial state is $|\psi^{2i}\rangle =|0\dots 00\rangle, |\psi^{2i+1}\rangle =|0\dots 01\rangle$, for $i\in\{0,\dots,k-1\}$, $|\psi^{2k}\rangle =|0\rangle$. So, for odd groups the last one  (with index $\lceil \log d \rceil$) is $|1\rangle$.

{\bf On the $r$-th layer} the program $P$ starts from the state $|\psi^{q}\rangle=|f^{(q)}(0)+ d(q \mbox{ }mod\mbox{ }2)\rangle$,  for $q\in\{1,\dots,r-1\}$. And other qubits were not changed. Let the $i$-th block contains variables $x^i_0,\dots,x^i_{\lceil \log_2 d  \rceil-1}$

The program applies the unitary matrix $D^{x^i_j}_j$ to $|\psi^{r-1}\rangle|\psi^{r}\rangle$ for $i\in\{0,\dots,2m-1\},j\in\{0,\dots,\lceil \log d \rceil-1\}$ on variable $x^i_j$. And does not change  other $|\psi^v\rangle$ for $v\neq r, r-1$. Using this matrices $P$ stores  $f^{(r)}(0)$ into qubits $|\psi^r\rangle$.

The $4d^2 \times 4d^2$ matrix $D^{p}_j$, for $p\in\{0,1\}$, and storing procedure are the following:

$D^p_j = \begin{pmatrix}
G_{0,j}^p & 0 & \cdots & 0 & 0 & 0 & \cdots & 0\\
0 & G_{1,j}^p & \cdots & 0 & 0 & 0 & \cdots & 0\\         
\vdots & \vdots & \ddots & \vdots & \vdots & \vdots & \ddots  & \vdots\\
0 & 0 & \cdots & G_{d-1,j}^p & 0 & 0 & \cdots & 0\\
0 & 0 & \cdots & 0 & I^{\otimes \lceil d + 1\rceil} & 0 & \cdots & 0\\
0 & 0 & \cdots & 0  & 0 & I^{\otimes \lceil d + 1\rceil} & \cdots & 0\\
\vdots & \vdots & \ddots & \vdots & \vdots & \vdots & \ddots  & \vdots\\
0 & 0 & \cdots & 0  & 0 & 0 & \cdots & I^{\otimes \lceil d + 1\rceil}
\end{pmatrix}$ for odd $r$ and

$D^p_j = \begin{pmatrix}
I^{\otimes \lceil d + 1\rceil}  & 0 & \cdots & 0 & 0 & 0 & \cdots & 0\\
0 & I^{\otimes \lceil d + 1\rceil}  & \cdots & 0 & 0 & 0 & \cdots & 0\\         
\vdots & \vdots & \ddots & \vdots & \vdots & \vdots & \ddots  & \vdots\\
0 & 0 & \cdots & I^{\otimes \lceil d + 1\rceil}  & 0 & 0 & \cdots & 0\\
0 & 0 & \cdots & 0 & G_{0,j}^p & 0 & \cdots & 0\\
0 & 0 & \cdots & 0  & 0 & G_{1,j}^p  & \cdots & 0\\
\vdots & \vdots & \ddots & \vdots & \vdots & \vdots & \ddots  & \vdots\\
0 & 0 & \cdots & 0  & 0 & 0 & \cdots & G_{d-1,j}^p
\end{pmatrix}$ for even $r$,

where $G_{i,j}^0$ and $G_{i,j}^1$ are $2d\times 2d$ matrices for storing of the variable $x^i_j$ into the $j$-th qubit of $|\psi^r\rangle$. And $G_{z,j}^0=G_{,j}^1=I^{\otimes \lceil \log d + 1\rceil}$, if $z\neq i$.

Storing procedure is the following:

Let us consider variables of the $i$-th block. We use $2\lceil \log d\rceil$ transition matrices. For the $j$-th bit of the block a pair of matrices is the following:
$G^0_{i,j}=I^{\otimes \lceil \log d + 1\rceil}$, $G^1_{i,j}$ is such that we apply $NOT$ gate for the $j$-th qubit of $|\psi^r\rangle$ and $I$ gate to all others.  $I$ and $NOT$ are $2\times 2$ matrices such that $I$ is a diagonal $1$-matrix and $NOT$ is an anti-diagonal $1$-matrix. So, qubits $|\psi^{r-1}\rangle$ determine which block will be stored on the layer.

Hence $f^r(0)$ is stored in the first $\lceil \log d\rceil$ qubits of $|\psi^r\rangle$. 
The last qubit shows us which function should be considered $f_A$ or $f_B$.

{\bf On the last layer} the program $P$ starts form the state $|\psi^{q}\rangle=|f^{(q)}(0)+ d(q \mbox{ }mod\mbox{ }2)\rangle$,  for $q\in\{1,\dots,2k-1\}$. And $|\psi^{2k}\rangle =|0\rangle$.

The program applies $U^{x^i_j}=D^{x^i_j}_0$ to $|\psi^{2k-1}\rangle|\psi^{2k}\rangle$ for $i\in\{0,\dots,2m-1\}$. Using $U^{x^i_j}$ we compute XOR of all qubits from the $(f^{(2k-1)}(0)+d)$-th block and store this result into $|\psi^{2k}\rangle$.

Finally, the program $P$  measures qubit $|\psi^{2k}\rangle$ and gets the right result with probability $1$.

By construction $P$ computes  $PJ_{2k-1,n}(X)$ and the program is commutative. Then we apply to the program the following theorem from \cite{kk2017} and get $2k$-QOBDD $P'$ of width $O(n^{2k})$ for xor-reordered version of $PJ_{2k-1,n}$, means that $XRPJ_{2k-1,n}(X)$.

\begin{theorem}[\cite{kk2017}]
Let the Boolean function $f$ over $X=(x_1, \cdots, x_n)$ be such that $N^{id}(f)\geq d(n)$ and a commutative k-QOBDD $P$ of width $g(n)$ computes $f$. Then there is total Boolean function $f'$, total xor-reordered version of $f$,  such that $N(f')\geq d(q)$, where  $n=q(\lceil\log q\rceil+1)$. And there is $k$-QOBDD $P'$ of width $g(q) \cdot q$ which computes $f'$. 
\end{theorem}

\section{Proof of Lemma \ref{lm:MXPJ-subfunc}}\label{apx:MXPJ-subfunc}

Let us consider the natural order  $id=(1,\dots,n)$ and the partition $\pi=(X_A,X_B)\in \Pi(\theta)$  such that $X_A=(x_1,\dots,x_{kd\lceil \log_2 d\rceil})$, $X_B=X/X_A$

For an input $\nu$ we have the partition $(\sigma,\gamma)$ with respect to
$\pi$. 

 Let $Val(X_A,j,i)$ be a value of a block, for $i \in \{0 \cdots k-1\}, j\in\{0,\dots,d-1\}$ such that $Val(X_A,j,i)=a_{i+1,j+1}$ from Figure \ref{fig:ab}. $Val(X_B,j,i)$ be a value of a block, for $i \in \{0 \cdots k-1\}, j\in\{d,\dots,2d-1\}$ such that $Val(X_B,j,i)=b_{i+1,j-d+1}$

We define the set $\Sigma\subset\{{0,1\}^{|X_A|}}$. The set satisfies the following conditions: for
$\sigma\in\Sigma$ we have

\begin{enumerate}
\item $Val(\sigma,0,t)=Val(\sigma,1,t)=Val(\sigma,2,t)=0$, for $0\leq t\leq k-1$;
\item $Val(\sigma,u,0)=0$, for $u\in\{0,\dots,d-1\}$;
\item $Val(\sigma,u,k-2)=Val(\sigma,u,k-1)=0$, for $u\in\{0,\dots,d-1\}$;
\item $Val(\sigma,u,t)\oplus Val(\sigma,u+\lfloor d/3-1 \rfloor,t) =1$,

 $Val(\sigma,u,t)\oplus Val(\sigma,u+2\lfloor d/3-1 \rfloor,t) =2$,
 
  for $0< t< k-2$ and $3\leq u\leq 2+\lfloor d/3-1 \rfloor$.

\end{enumerate}

Let us consider two different inputs $\sigma,\sigma'\in
\Sigma$ and corresponding mappings $\tau$ and $\tau'$. Let us show
that subfunctions $MXPJ_{2k,d}|_\tau$ and $MXPJ_{2k,d}|_{\tau'}$ are
different functions. The inequality of $\sigma$ and $\sigma'$ means that there are $r\in\{1,\dots,k-3\}$ and $z\in\{3,\dots, 2+\lfloor d/3-1\rfloor\}$ 
such that $s'=Val(\sigma',z,r)\neq Val(\sigma,z,r)=s$. Let us choose
$\gamma\in \{0,1\}^{|X_B|}$ such that:
\begin{enumerate}
\item $Val(\gamma,d+0,t)=0$, for $0\leq t < r-1$;
\item $Val(\gamma,d+0,r-1)=z$;
\item $Val(\gamma,d+s,r)=z\oplus(z+\lfloor d/3-1 \rfloor)$;
\item $Val(\gamma,d+s',r)=z\oplus(z+2\lfloor d/3-1 \rfloor)$;
\item $Val(\gamma,d+1,r+1)=1\oplus(z+\lfloor d/3-1 \rfloor)$ and $ Val(\gamma,d+2,r+1)=2\oplus(z+2\lfloor d/3-1 \rfloor)$;
\item $Val(\gamma,d+1,t)=Val(\gamma,d+2,t)=0$, for $r+2\leq t\leq k-2$;
\item $Val(\gamma,d+1,k-1)=1$, $Val(\gamma,d+2,k-1)=0$;
\end{enumerate}

Let $\nu=(\sigma,\gamma)$ and $\nu'=(\sigma',\gamma)$. 

Because of properties 1 of $\Sigma$ and $\gamma$, we have $f^{(j)}(\nu,t)=f^{(j)}(\nu',t)=0$, for $0\leq t\leq 2r-1$.

Due to property 2 of $\gamma$, we have $f^{(2r)}(\nu)=f^{(2r)}(\nu')=z$. Then  $f^{(2r+1)}(\nu)=s$ and $f^{(2r+1)}(\nu')=s'$.

Because of properties 3 and 4 of $\gamma$, we have $f^{(2r+2)}(\nu)=z+\lfloor d/3-1 \rfloor$ and $f^{(2r+2)}(\nu')=z+2\lfloor d/3-1 \rfloor$.

Due to property 4 of $\Sigma$, we have $f^{(2r+3)}(\nu)=1$ and $f^{(2r+3)}(\nu')=2$.

Because of property 5 of $\gamma$, we have $f^{(2r+4)}(\nu)=1$ and $f^{(2r+4)}(\nu')=2$.

Due to property 1 of $\Sigma$ and  property 6 of $\gamma$, we have $f^{(t)}(\nu)=1$, $f^{(t)}(\nu')=2$ for $2r+4\geq t \geq 2k-1$.

Then because of property 7 of $\gamma$, we have 
$MXPJ_{2k,d}(\sigma,\gamma)=1$, $MXPJ_{2k,d}(\sigma',\gamma)=0$. Therefore, $MXPJ_{2k,d}|_\tau(\gamma)\neq MXPJ_{2k,d}|_{\tau'}(\gamma)$ and
$MXPJ_{2k,d}|_\tau\neq MXPJ_{2k,d}|_{\tau'}$.

Let us compute $|\Sigma|$. Because of properties of $\Sigma$, we have $|\Sigma|\geq d^{\lfloor d/3-1 \rfloor (k-3)}$. Therefore, $N^{\pi}(MXPJ_{2k,d})\geq d^{\lfloor d/3-1 \rfloor (k-3)}$ and due to definition of $N^{id}$, we have $N^{id}(MXPJ_{2k,d})\geq d^{\lfloor d/3-1 \rfloor (k-3)}$.
\Endproof

\section{Proof of Theorem \ref{th:hierarchy-big}}\label{apx:hierarchy-big}
By the definition, we have BQP$_{1/8}$-$k$OBDD$\subseteq$BQP$_{1/8}$-$(2k)$OBDD. Let us prove inequality of these classes.  
Let us consider $XRPJ_{2k-1,n}$. Due to Lemma \ref{lm:pj-kobdd}, each $k$-QOBDD computing the function has size:  
  
$2^{\Omega( n/(k^3\log n)  - \log(n/\log n))}\geq$ $  
2^{\Omega(n/(n\log^{-3} n \log n)- \log (n/\log n))}
=n^{\Omega(\log n)}$.

Therefore, the program has more than a polynomial size. Hence $XRPJ_{2k-1,n}\not\in$ BQP$_{1/8}$-$k$OBDD and $XRPJ_{2k-1,n}\in$ BQP$_{1/8}$-$(2k)$OBDD,  
  due to the second claim of Lemma \ref{lm:pj-kobdd}.  

\end{document}